\documentclass[a4paper,11pt]{article}
\usepackage{pos}
\usepackage{booktabs}
\usepackage{hyperref}
\usepackage[utf8]{inputenc}

\title{\textit{SkyNET-scape Room}: An Escape Room to Explore Astroparticle Physics}
\ShortTitle{\textit{SkyNET-scape Room}: An Escape Room to Explore Astroparticle Physics}

\author*[a,b]{E. Prandini}
\author[b]{S. Hemmer}
\author[c]{I. Viale}
\author[d]{M. Branca}
\author[a]{L. Campagnoni}
\author[e]{G. Cataldi}
\author[f]{M. Circella}
\author[a]{J.-P. Jonckheere}
\author[g]{E. Leonora}
\author[a,b]{G. Silvestri}

\affiliation[a]{Padova University, DFA, I-35131, Italy}
\affiliation[b]{INFN, Sezione di Padova, I-35131 Padova, Italy}
\affiliation[c]{INFN, Sezione di Torino, I-10125 Torino, Italy}
\affiliation[d]{Liceo "Quinto Ennio" Gallipoli, I-73014 Gallipoli (Lecce), Italy}
\affiliation[e]{INFN, Sezione di Lecce, I-73100  Lecce, Italy}
\affiliation[f]{INFN, Sezione di Bari, I-70121 Bari, Italy}
\affiliation[g]{INFN, Sezione di Catania, I-95123 Catania, Italy}

\emailAdd{elisa.prandini@unipd.it}
\emailAdd{sabine.hemmer@pd.infn.it}

\abstract{Innovative science communication is key to engaging the public with complex topics such as astroparticle physics. As part of the Italian PRIN 2022 funding initiative, we are developing SkyNET-scape Room, an interactive escape room designed to introduce participants to the main messengers of the high-energy universe, namely cosmic rays, gamma rays, and neutrinos.

The experience is structured into three interactive stations, each focusing on a specific messenger and its detection method. Small teams of visitors will collaborate to solve puzzles and quizzes related to gamma-ray observations, neutrino detection, and cosmic-ray composition. Using a combination of audio, video, and written materials, participants will engage with real scientific concepts while working together to "escape." The ultimate goal is to spark curiosity, teamwork, and excitement for science, making astroparticle physics more accessible and engaging.

In this contribution, we will
present the design, implementation, and preliminary feedback from test
sessions, emphasizing how gamification and interactive storytelling can
serve as powerful tools for public engagement.}

\ConferenceLogo{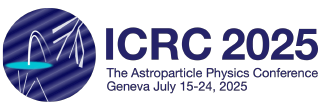}

\FullConference{39th International Cosmic Ray Conference (ICRC2025)\\
 15–24 July 2025\\
Geneva, Switzerland\\}

\begin{document}
\maketitle

\section{Introduction: Escape Room as an Innovative Outreach Initiative}

Escape rooms are immersive, team-based experiences where participants solve a series of puzzles and riddles within a limited time to "escape" from a scenario. Originally developed as entertainment, escape rooms have evolved into educational tools thanks to their ability to combine problem-solving with storytelling and collaboration \cite{2021Front...622860}. Their format naturally encourages critical thinking, communication, and engagement, making them a compelling medium for public engagement.

The format’s success in science communication has been demonstrated by initiatives such as HEPscape! \cite{CAVALLARI:2025W1}, developed in 2021 by a team from INFN. HEPscape! invites participants into a simulated control room of the Large Hadron Collider (LHC) at CERN. The experience uses projectors, lighting effects, and physical clues to create an immersive environment where players solve puzzles that introduce concepts from high-energy physics. The escape room format allows participants to learn about particle accelerators and detector technologies in an interactive and enjoyable way. Importantly, HEPscape! is portable and can be set up in approximately three hours, making it well-suited for science fairs and outreach events. Its materials are low-cost and easy to replicate, and the format has been successfully translated into different languages and adapted for various age groups. Feedback from events where HEPscape! was deployed highlights its effectiveness in sparking curiosity and facilitating informal learning.

\subsection{The PRIN 2022 Initiative and Our Objectives}
Our project, SkyNET-scape Room, is part of the SkyNET--Deep Learning for Astroparticle Physics project, developed within the framework of the PRIN 2022 (Progetti di Rilevante Interesse Nazionale) initiative, funded by the Italian Ministry of University and Research (MUR). PRIN supports fundamental research projects with national relevance, promoting interdisciplinary collaboration and public engagement. In our case, this funding allows us to merge scientific research in astroparticle physics with innovative science communication strategies.

The primary objective of SkyNET-scape Room is to design and implement a portable escape room experience tailored to public science events such as the European Researchers’ Night, which occurs annually in late September. Through this experience, we aim to introduce participants to the main messengers of the high-energy universe—gamma rays, neutrinos, and cosmic rays—by combining interactive storytelling with hands-on puzzle-solving. Our escape room is being designed with mobility, scalability, and accessibility in mind, ensuring that it can be easily transported, assembled, and adapted to different contexts and audiences.

This project is carried out in collaboration with OCRA (Outreach Cosmic Ray Activities) \cite{Hemmer:2021oe,Aramo:20195l}, born in 2018 as a national framework project of INFN for the numerous public engagement activities in the field of cosmic ray physics.


\section{Design Phase of the \textit{SkyNET-scape Room}}

The design of the \textit{SkyNET-scape Room} began in late 2023, following the start of the PRIN 2022 funding period. A small team of three members was formed, which started holding regular monthly meetings to brainstorm and develop the core ideas of the project.

After initial discussions, the team converged on the concept of an escape room structured around three interactive stations, each corresponding to a different cosmic messenger. The choice of detectors was guided by our collaboration with OCRA, drawing from active INFN experiments: Pierre Auger Observatory for charged cosmic rays, MAGIC/CTAO for gamma rays, and KM3NeT for neutrinos.

To better understand the challenges of such an endeavor, we consulted with colleagues involved in the \textit{HEPscape!} room. These discussions helped us identify both the strengths and weaknesses of their setup, which informed several of our design choices.

Before advancing to the physical development phase, we defined the main framework and constraints of the project. The target audience was set to participants aged 12 and above, ensuring cognitive engagement without oversimplifying the content. The educational goals were threefold: (i) to spark interest in astroparticle physics, (ii) to encourage cooperation and teamwork, reflecting the collaborative nature of scientific research, and (iii) to introduce participants to hands-on, experiment-inspired activities, particularly in detection techniques.

Finally, the escape room was designed with a clear set of practical requirements: it must be portable, easy to assemble, and manageable by only one or two operators. Materials are chosen to be commercially available, and all content is closely tied to real experiments to ensure scientific relevance and authenticity.

\section{Implementation}

The implementation of the \textit{SkyNET-scape Room} began approximately in Summer 2024, following the conclusion of the design phase. As the project moved forward, flexibility and practicality became central considerations in shaping the experience.

Depending on the type of event and the number of participants, the activity can be conducted in two main formats: either sequentially, where all participants proceed through each station one after the other, or in parallel, where the group is divided into three smaller teams, each tackling a different station simultaneously. In both cases, the activity is hosted within a single physical room, carefully arranged to support the modular layout of the stations.

The experience opens with a short introductory narrative lasting about three minutes. This prologue is delivered through a recorded voice accompanied by projected images on a large screen. The introduction sets an immersive and evocative tone: an enigmatic astrophysical event that occurred millions of years ago is now reaching Earth. All astronomical observatories are on alert, and the participants are called upon to help determine the origin of the event, which involves three different cosmic messengers—charged cosmic rays, gamma rays, and neutrinos.

Each station is centered on one of these messengers and is associated with a relevant detector: the Pierre Auger Observatory for cosmic rays, the MAGIC/CTAO telescopes for gamma rays, and KM3NeT for neutrinos. Completing the challenges at each station reveals a part of the celestial coordinates. When all three parts are collected, the full coordinates can be used to identify the location of the mysterious event.

Each station features three challenges. Among these, at least one is designed to involve experimental or hands-on reasoning. For example, participants may be asked to determine the curvature radius of a concave mirror, as used in imaging atmospheric Cherenkov telescopes, or estimate the water volume contained in a surface detector tank like those at the Pierre Auger Observatory. Other tasks are more playful and accessible, such as revealing a hidden word buried in a box filled with sand, mimicking the Mediterranean seabed, in reference to the KM3NeT neutrino telescope.

The physical setup of the room includes three tables, each equipped with models of the respective instruments, posters, and small furnishing elements to enhance the immersive atmosphere. Up to three projectors may be used to create visual effects or support interactive content during the activity.

During the 2024/2025 school year, we also collaborated with a secondary school in Gallipoli (Lecce), where a group of students developed a custom card-reading system based on Arduino. This system will be integrated into one of the room's stations to introduce an additional level of interactivity and realism. This collaboration provided an opportunity to connect formal education with public engagement, and encouraged student engagement with applied physics and engineering.

Once all stations are completed and the full set of coordinates is obtained, participants are invited to input them into the planetarium software Stellarium, where the sky position of the event is revealed. This marks the conclusion of the experience with a scientifically meaningful outcome.

\subsection{Feedback during the International Cosmic Day 2024}
A first test of the activity was carried out in November 2024 with students participating in the International Cosmic Day in Padua, Italy. During the event, we presented the concept of the escape room and invited the students to try out some of the puzzles related to the gamma-ray station, which was the first to be fully developed. This early testing phase allowed us to identify certain issues with the proposed measurements and to collect valuable feedback and suggestions from the participants. Several of these suggestions were taken into account and helped improve the overall design of the activity.

\section{Conclusions and Outlook}

The SkyNET-scape Room project is currently in its full development and fine-tuning phase. After completing the design and initial testing stages, we are now focusing on optimizing the structure, refining the puzzles, and finalizing the full three-station setup. 

We already have two important public engagement events ahead of us, where the escape room will be presented to the general public. The first is the European Researchers' Night, scheduled for the end of September 2025 in Padua\footnote{\url{https://science4all.it}}. The second is the Science Festival in Schio, Vicenza, planned for October 2025\footnote{\url{https://www.fesav.it/}}. These events will serve as key opportunities to assess the impact of the activity, engage a wide and diverse audience, and gather valuable feedback to improve and refine the escape room for future science communication initiatives.

\acknowledgments
Elisa Prandini, Sabine Hemmer, and Ilaria Viale acknowledge funding for the project “SKYNET: Deep Learning for Astroparticle Physics”, 693 PRIN 2022 (CUP: D53D23002610006).

Moreover, we are grateful to: 
Lorenzo Caccianiga and Lorenzo Ramella for the Auger Map; Anna Dalla Bona for the realization of the movie; Loris Ramina and David Agguiaro for the unvaluable technical help; Marina Manganaro, Serena Loporchio and the MAGIC outreach working group for the 3D MAGIC model;  the students of the Liceo "Quinto Ennio", based in Gallipoli, Lecce (Italy) for the development of the arduino-based card reader;  the students participating to the International Cosmic Day, Padua (2024), for the useful suggestions; and Michele Doro and Andrea Rossi for their help.

\begin{small}

\end{small}
\end{document}